# Cavity-assisted enhanced and dephasing immune squeezing in the resonance fluorescence of a single quantum dot

Parvendra Kumar[1,2,*] and Agnikumar G. Vedeshwar[1,†]

[1]*Thin Film Laboratory, Department of Physics and Astrophysics, University of Delhi, Delhi 110007, India*
[2]*Department of Physics, Bhagini Nivedita College, University of Delhi, New Delhi-110043, India*
[*]parvendra1986@gmail.com, [†]agni@physics.du.ac.in

We theoretically demonstrate the enhanced and dephasing immune squeezing in the resonance fluorescence of a single quantum dot (QD) confined to a pillar-microcavity and driven by a continuous wave laser. We employ a formalism based on Polaron master equation theory for incorporating the influence of exciton-phonon coupling quite accurately in the dot-cavity system. We show a significant enhancement of squeezing due to cavity coupling of the QD as compared to that of an ideal single two-level system in free space. Particularly, we show a four-fold enhancement in squeezing as compared to that of a single QD without cavity coupling. We further demonstrate the persistence of squeezing even when the pure dephasing becomes greater than the radiative decay rate. These novel features are attributed to the cavity-enhanced coherence causing partial reduction of the deteriorating effects of phonon-induced incoherent rates. We also show that the deteriorating effects of phonon-induced incoherent rates on squeezing can be partially circumvented by properly adjusting the detunings.

## I. INTRODUCTION

The system of a single quantum emitter coupled to the tailored electromagnetic vacuum modes is of both fundamental and practical interest due to its novel spectral features. Tailoring of the ambient electromagnetic modes of a quantum emitter, by placing it inside a cavity, give rise to several practically useful effects such as inhibition and acceleration of natural decay rate of the quantum emitter, vacuum Rabi splitting, and Mollow triplet [1-4]. Further, fundamental understanding of such systems have led to the observation of several intriguing phenomena such as photon antibunching, entangled photons, and squeezed light. These are now serving as the key building blocks in the development of quantum technologies ranging from quantum information to quantum metrology [5-10]. Particularly, squeezed light is quite useful and essential in quantum metrology for carrying out the measurements beyond the standard quantum limit [11-13]. Squeezing refers to a possible reduction in the fluctuations of one of the canonically conjugate variables below the minimum at the expense of enhancing that of the other within the Heisenberg's uncertainty principle. Squeezed light generation have been demonstrated through quadratic process and Kerr effect in nonlinear systems [14–15], four-wave mixing in atomic and solid states systems [16–19], cavity-QED [20, 21], Bose–Einstein condensation [22], and optomechanical systems [23–26]. All these processes depend upon the quadratic form of the bosonic operators and multi-photon processes.

It was first predicted in 1981 and later realized experimentally that squeezing can be obtained through a radically different approach that does not require quadratic form of the bosonic operators [27]. It involves the interaction of two-level emitter with a resonant light field. This unique form of squeezing stems from a build-up and survival of steady state coherence in the weak excitation regime. Recently such form of squeezing was observed both experimentally and theoretically in the resonance fluorescence from a resonantly excited quantum dot (QD) [28, 29]. The achievable degree of squeezing was found to be quite low because of the limited coherence generation in the weak excitation regime. Additionally, unlike the real atoms, QDs are unavoidably coupled to the phonon bath of lattice vibrations, which generally introduces the phonon-induced dephasing and incoherent scattering of the exciton states of the QDs [30-36]. These phonon-induced incoherent processes were also shown to greatly limit the achievable degree of squeezing by hindering the building-up of exciton coherence. Furthermore, there was no squeezing obtained whenever the pure dephasing rate of QD exciton state becomes comparable or greater than the natural decay rate, which is a quite common scenario in QDs [28]. However, there exist few proposals for enhancing the squeezing either by advantageously harnessing the exciton-phonon coupling in a very strongly driven single QD [37] or by emitter-cavity coupling [38].

In this paper, we show that the coupling of a cavity mode to an appropriately driven QD can facilitate the enhanced and dephasing immune squeezing. Numerically calculated value of squeezing is found to be much greater than the maximum obtainable squeezing from an ideal two-level system in free space. We also show a four-fold enhancement in squeezing as compared to that is obtained in a single QD without cavity coupling. It is also shown that a fair amount of squeezing persists even when the pure dephasing rate exceeds the natural decay rate. The enhanced and dephasing immune squeezing is shown to be facilitated by the cavity-enhanced coherence. It is worthy to mention that we also utilize the cavity coupling advantageously for obtaining the enhanced squeezing similar to the Ref. [38]. However, in contrast, we employ a rigorous theoretical formalism for incorporating the unavoidable exciton-phonon coupling in a realistic dot-cavity system at the typical cryogenic temperatures. We further derive a simple and effective polaron master equation with the analytical forms of various phonon-induced incoherent rates. This enabled us to explicitly analyze and in fact to partially reduce the deteriorating effects of phonon-induced incoherent rates on squeezing by properly adjusting the detunings.

## II. THEORETICAL MODEL
### A. Model Hamiltonian and Polaron Transformation

We consider a single self-assembled InGaAs/GaAs QD embedded in a single mode pillar-microcavity. Single QD can be modeled as a two-level system with ground and exciton states interacting with single cavity mode driven by a continuous wave (CW) laser field. The unavoidable coupling of exciton state to the longitudinal acoustic (LA) phonons is included in our theoretical model for simulating realistic CW laser driven QD cavity system quite accurately. The Hamiltonian of the model system describing the interaction of a QD with LA phonons with the cavity mode excited by CW laser field can be written as,

$$H = \hbar\Delta_{xl}\sigma^+\sigma^- + \hbar\Delta_{cl}a^\dagger a + \frac{\hbar\Omega}{2}(\sigma^+ + \sigma^-) + \hbar g(\sigma^+ a + a^\dagger \sigma^-) + \sum_q \hbar\omega_q b_q^\dagger b_q + \sigma^+\sigma^- \sum_q \hbar\lambda_q(b_q^\dagger + b_q), \quad (1)$$

where $\Delta_{xl} = \omega_x - \omega_l$ represents the detuning of CW laser field with respect to exciton state, $\Delta_{cl} = \omega_c - \omega_l$ is the detuning of laser field with respect to cavity mode, $\sigma^+$ and $\sigma^-$ describe the creation and annihilation of the exciton state, $\Omega = \mu E/\hbar$ is the Rabi frequency which quantifies the coupling strength of CW laser with exciton, $g$ is the dot-cavity coupling strength, $b_q^\dagger$ and $b_q$ represent the creation and annihilation operators for mode $q$ of phonon bath, $\lambda_q$ represents the exciton-phonon coupling by means of deformation potential.

The exciton-phonon interaction in terms of effective polaron renormalized QD-laser and QD-cavity couplings can be realized by transforming the Hamiltonian (Eq. (1)) to polaron frame as $H' = e^P H e^{-P}$, where $P = \sigma^+\sigma^- \sum_q \lambda_q/\omega_q (b_q^\dagger - b_q)$ [39, 40]. The polaron-transformed Hamiltonian in terms of modified QD-laser system, phonon bath, and interaction part respectively are now given by,

$$H'_{sys} = \hbar(\Delta_{xl} - \Delta_P)\sigma^+\sigma^- + \hbar\Delta_{cl}a^\dagger a + \langle B \rangle X_g, \quad (2a)$$

$$H'_B = \sum_q \hbar\omega_q b_q^\dagger b_q, \quad (2b)$$

$$H'_I = X_g\zeta_g + X_u\zeta_u, \tag{2c}$$

where, $X_g = \frac{\hbar\Omega}{2}(\sigma^+ + \sigma^-) + \hbar g(\sigma^+ a + a^\dagger\sigma^-)$, $X_u = \frac{i\hbar\Omega}{2}(\sigma^+ - \sigma^-) + i\hbar g(\sigma^+ a - a^\dagger\sigma^-)$. Further, $\zeta_g$ and $\zeta_u$ represent the fluctuations operators defined as: $\zeta_g = \frac{1}{2}(B_+ + B_- - 2\langle B\rangle)$ and $\zeta_u = \frac{1}{2}(B_+ - B_-)$. The coherent displacement operators of phonon modes, $B_\pm$ will then be $B_\pm = exp\left[\pm\sum_q \frac{\lambda_q}{\omega_q}(b_q^\dagger - b_q)\right]$. The Hamiltonian, $H'_I$, describes the interaction of phonon-induced fluctuations with the CW laser driven quantum dot. The thermally averaged phonon displacement operators has the form $\langle B_+\rangle = \langle B_-\rangle = \langle B\rangle = exp\left[-\frac{1}{2}\int_0^\infty d\omega \frac{J(\omega)}{\omega^2}coth\left(\frac{\hbar\omega}{2K_BT}\right)\right]$, where $T$ represents the temperature of phonon-bath. The polaron shift $\Delta_P$ is given by $\Delta_P = \int_0^\infty d\omega \frac{J(\omega)}{\omega}$. We further simplify the formulation by including the polaron shift in the definition of $\omega_x$ itself by redefining the detuning as $\Delta_{xl} = \Delta_{xl} - \Delta_P$. The phonon spectral function, $j(\omega)$, characterizing the exciton-phonon coupling can be expressed as $j(\omega) = \alpha_P\omega^3 exp\left(-\frac{\omega^2}{2\omega_b^2}\right)$, where $\alpha_P$ represents the strength of exciton-phonon coupling and $\omega_b$ represents the phonon cutoff frequency.

**B. Polaron Master Equation Theory**

Several distinct theoretical approaches such as variational formulation, quasi-adiabatic path integral, and polaron transformation techniques have been employed for incorporating the influence of exciton-phonon interaction on the dynamics of driven QDs [41-43]. However, master equation based on the polaron transformation technique is quite well-known for facilitating the accurate results and much faster computation under the suitable parameters regime: $(\Omega/\omega_b)^2(1 - \langle B\rangle^4) \ll 1$ [44]. Therefore in this work, we employ the polaron master equation theory for investigating the influence of exciton-phonon interaction on squeezing in the driven QD-cavity system. Following the procedural details given in Refs. [42-45], we derive time-local master equation (ME) of reduced density operator, $\rho(t)$, of the QD-cavity laser-phonon system specifically considered in this paper. The Polaron ME obtained under second-order Born approximation of QD-phonon coupling reads as

$$\frac{d\rho(t)}{dt} = -\frac{i}{\hbar}[H'_{sys},\rho(t)] - \frac{1}{\hbar^2}\int_0^\infty d\tau \sum_{m=g,u}\{G_m(\tau)[X_m, e^{-iH'_{sys}\tau}X_m e^{iH'_{sys}\tau}\rho(t)] + H.C.\} + L[\rho(t)], \tag{3}$$

Where polaron Green functions, $G_m(\tau) = \langle\zeta_m(\tau)\zeta_m(0)\rangle$, are calculated to be $G_g(\tau) = \langle B\rangle^2\{cosh[\phi(\tau)] - 1\}$ and $G_g(\tau) = \langle B\rangle^2 sinh[\phi(\tau)]$ [39]. Polaron correlation function, $\phi(\tau)$, reads as $\phi(\tau) = \int_0^\infty d\omega \frac{j(\omega)}{\omega^2}\left[coth\left(\frac{\hbar\omega}{2K_BT}\right)cos(\omega\tau) - isin(\omega\tau)\right]$. The second term represents the phonon-induced incoherent processes. This term explicitly depends on the Phonon-bath temperature and Rabi frequency via $G_m(\tau)$ and $X_m$, respectively. The superoperator term, $L[\rho(t)] = \frac{\gamma}{2}£[\sigma^-]\rho(t) + \frac{\gamma'}{2}£[\sigma^+\sigma^-]\rho(t) + \frac{\kappa}{2}£[a]\rho(t)$, is added phenomenologically for incorporating the radiative ($\gamma$) and pure dephasing ($\gamma'$) rates of exciton state along with the cavity decay ($\kappa$) rate. For gaining the essential physical insight into the phonon-induced processes, we simplify the Eq. [3] by approximating, $e^{-iH'_{sys}\tau}X_m e^{iH'_{sys}\tau} \approx e^{-iH'_0\tau}X_m e^{iH'_0\tau}$ with $H'_{sys} = \hbar\Delta_{xl}\sigma^+\sigma^- + \hbar\Delta_{cl}a^\dagger a$. This approximation gives the accurate and much faster results only when $\Omega^{-1}$ and $g^{-1}$ are much larger than the phonon correlation time ($\tau_{ph} \approx 2\ ps$) or when the detunings ($\Delta_{xl}, \Delta_{cx}$) are much larger than $\Omega$ and $g$ [32]. The simplified effective polaron master equation now reads as

$$\frac{d\rho(t)}{dt} =$$

$$-\frac{i}{\hbar}[H'_{sys}, \rho(t)] + \frac{\Gamma_{ph}^{\sigma^+}}{2}\mathcal{L}[\sigma^+]\rho(t) + \frac{\Gamma_{ph}^{\sigma^-}}{2}\mathcal{L}[\sigma^-]\rho(t) + \frac{\Gamma_{ph}^{a^\dagger\sigma^-}}{2}\mathcal{L}[a^\dagger\sigma^-]\rho(t) + \frac{\Gamma_{ph}^{\sigma^+a}}{2}\mathcal{L}[\sigma^+a]\rho(t) + L[\rho(t)], \qquad (4)$$

where, $\Gamma_{ph}^{\sigma^+}$ ($\Gamma_{ph}^{\sigma^-}$) represents the phonon-induced rate of incoherent excitation (rate of incoherent de-excitation) of QD exciton state and $\Gamma_{ph}^{a^\dagger\sigma^-}$ ($\Gamma_{ph}^{\sigma^+a}$) represents the phonon-induced rate of cavity photon creation accompanied by the decay of the exciton state (rate of cavity photon annihilation accompanied by the excitation of the exciton state). These phonon-induced incoherent rates are given by

$$\Gamma_{ph}^{\sigma^+/\sigma^-} = \frac{\Omega_R^2}{2}Re\left[\int_0^\infty d\tau e^{\mp i\Delta_{xl}\tau}(e^{\phi(\tau)} - 1)\right] \qquad 5(a)$$

$$\Gamma_{ph}^{\sigma^+a/a^\dagger\sigma^-} = 2g_R^2 Re\left[\int_0^\infty d\tau e^{\pm i\Delta_{cx}\tau}(e^{\phi(\tau)} - 1)\right], \qquad 5(b)$$

where, $\Omega_R = \langle B \rangle \Omega$, $g_R = \langle B \rangle g$, and $\Delta_{cx} = \omega_c - \omega_x$ are the phonon-renormalized Rabi frequency, phonon-renormalized cavity coupling strength, and the detuning of exciton state with respect to cavity mode, respectively. The value of mean phonon displacement, $\langle B \rangle$, is 0.91 for temperature $T = 4\,K$.

## C. Squeezing in the Resonance Fluorescence

Following Ref. [25, 35], the amplitude quadrature of fluorescence field is defined as: $E_\theta = (E^+ e^{+i\theta} + E^- e^{-i\theta})$, where $E^+$ and $E^-$ represent the positive and negative frequency components of the electric field, and $\theta$ represents the quadrature phase. The squeezing properties can be investigated by analyzing the normally ordered variance, $\langle:\Delta E_\theta^2:\rangle = \langle:E_\theta^2:\rangle - \langle E_\theta \rangle^2$, of the electric field quadrature as defined above. We use the correspondence between atomic and field operators, $E^+ = |\epsilon|\sigma^-$ and $E^- = |\epsilon|\sigma^+$, to calculate the normally ordered variance of electric field quadrature. The normally ordered variance is given by $\langle:\Delta E_\theta^2:\rangle = |\epsilon|^2(2(\langle\sigma^+\sigma^-\rangle - |\langle\sigma^+\rangle|^2) - 2|\langle\sigma^+\rangle|^2 \cos(2\theta))$ [26]. Assuming the proportionality constant, $|\epsilon|$, to be unity, we find that the quadrature variance is minimum for quadrature phase, $\theta = 0$ and is given by

$$\langle:\Delta E^2:\rangle = 2[\langle\sigma^+\sigma^-\rangle - 2|\langle\sigma^-\rangle|^2] \qquad (6)$$

The fluorescence field is squeezed if normally ordered variance is negative, corresponding to a noise reduction below the vacuum level. Thus the greater negative values of normally ordered variance are equivalent to the higher squeezing. Theoretically, the maximum negative value of -0.125 of normally ordered field variance can be obtained in the fluorescence of an ideal single two-level system in free space [27, 38]. For the calculation of exciton state population, $\langle\sigma^+\sigma^-\rangle$, and exciton coherence, $\langle\sigma^-\rangle$, we chose to use the effective polaron ME [Eq. (4)] due to its ability for facilitating the efficient and much faster computation. Of course, we verified the numerical results obtained by Eq. 3 and Eq. 4 and these are found to be matching quite well across the range of chosen parameters. Note that in addition to the satisfaction of validity range of polaron ME theory ( Eq. 3), $(\Omega/\omega_b)^2(1 - \langle B \rangle^4) \ll 1$, the validity condition of effective polaron master equation (Eq. 4), $\Omega^{-1}, g^{-1} \gg \tau_{ph}$, is also very well satisfied for the presently employed typical parameters, viz., $(\Omega/\omega_b)^2(1 - \langle B \rangle^4) = 0.015$ and $\Omega^{-1}, g^{-1} = 18.79\,ps$ for $\Omega, g = 220\,\mu eV$ at $T = 4K$. Therefore, the validity conditions for both full and effective polaron master equations are clearly satisfied.

## III. RESULTS AND DISCUSSIONS

For the investigation of squeezing in resonance fluorescence, we use the relevant parameters of InAs/GaAs self-assembled QDs for the simulation. The typical values of the chosen parameters are $\gamma = 2\,\mu eV$, $\alpha_P/(2\pi)^2 = 0.06\,ps^2$, $\gamma' = 0.5\,\mu eV$ and $\omega_b = 1\,meV$ [35, 36]. However, the values of decay and pure dephasing rates can vary in the range of $1 - 3\,\mu eV$ and $0.5 - 3\,\mu eV$ respectively for different dots depending on their varying size and shapes. Therefore, for the greater generality of our results, we also investigate the effects of variation in decay and dephasing rates by treating them also as variables. The other simulation parameters relating to the laser field and cavity are mentioned at appropriate places either in text or in the figure captions.

### A. Detuning Dependent Profile of Phonon-induced Excitation and De-excitation Rates

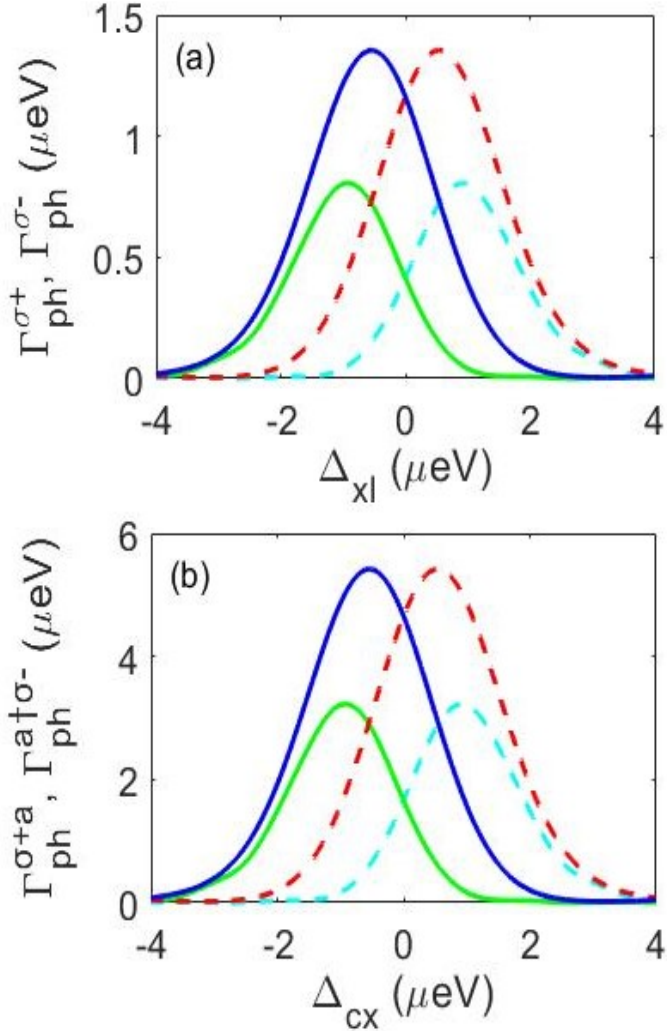

Fig. 1(color online) Detuning dependent profiles of phonon-induced excitation and de-excitation rates for a fixed value of Rabi frequency, $\Omega_R = 100\,\mu eV$, and cavity coupling strength, $g_R = 100\,\mu eV$ for the two cases:(a) $\Gamma_{ph}^{\sigma^-}$ (dashed cyan line), $\Gamma_{ph}^{\sigma^+}$ (solid green line) at $T = 4\,K$, and $\Gamma_{ph}^{\sigma^-}$ (dashed red line), $\Gamma_{ph}^{\sigma^+}$ (solid blue line) at $T = 10\,K$ and (b) $\Gamma_{ph}^{\sigma^+ a}$ (dashed cyan line), $\Gamma_{ph}^{a^\dagger \sigma^-}$ (solid green line) at $T = 4\,K$ and $\Gamma_{ph}^{\sigma^+ a}$ (dashed red line), $\Gamma_{ph}^{a^\dagger \sigma^-}$ (solid blue line) at $T = 10\,K$.

Firstly, we show in Fig. 1, the detuning dependent profiles of phonon-induced rates for the fixed values of Rabi frequency, $\Omega_R = 100\,\mu eV$, and cavity coupling strength, $g_R$ for two different temperatures, $T = 4\ and\ 10\ K$. The key feature of the figure 3(a) is the asymmetry between $I_{ph}^{\sigma^+}$ and $I_{ph}^{\sigma^-}$ as a function of the exciton-laser detuning, which is more pronounced at lower temperature of $4\ K$ than at $10\ K$. This is particularly due to the asymmetry between the $I_{ph}^{\sigma^+}(or I_{ph}^{\sigma^-})$ at the two temperatures. Note that $I_{ph}^{\sigma^+}$ is maximum at the point where $I_{ph}^{\sigma^-}$ almost vanishes at $\Delta_{xl} \approx -1meV$ and vice versa at $\Delta_{xl} \approx 1meV$. Similarly in Fig. 1(b), the detuning dependent profiles of $I_{ph}^{a^\dagger \sigma^-}$ and $I_{ph}^{\sigma^+ a}$ exactly follow the same trend as those of $I_{ph}^{\sigma^-}$ and $I_{ph}^{\sigma^+}$, but with the appropriate parameter, the cavity-exciton detuning, $\Delta_{cx}$. In our present work, we appropriately exploit this asymmetry as depicted in Fig. 3(b) to enhance the obtainable squeezing.

### B. Evolution of the Variance as a Function of Cavity-Laser Detuning

Further, we investigate the evolution of the variance of a laser driven QD as a function of cavity detuning, $\Delta_{cl}$ at three different Rabi frequencies, $\Omega_R$ without incorporating the exciton-phonon coupling by choosing phonon-bath temperature, $T = 0\ K$ as depicted in Fig. 2. We deliberately chose a large exciton-laser detuning, $\Delta_{xl} = \Omega_R$, for restricting the significant flow of population into the exciton state for achieving the maximally possible squeezing [see Eq. (6)].

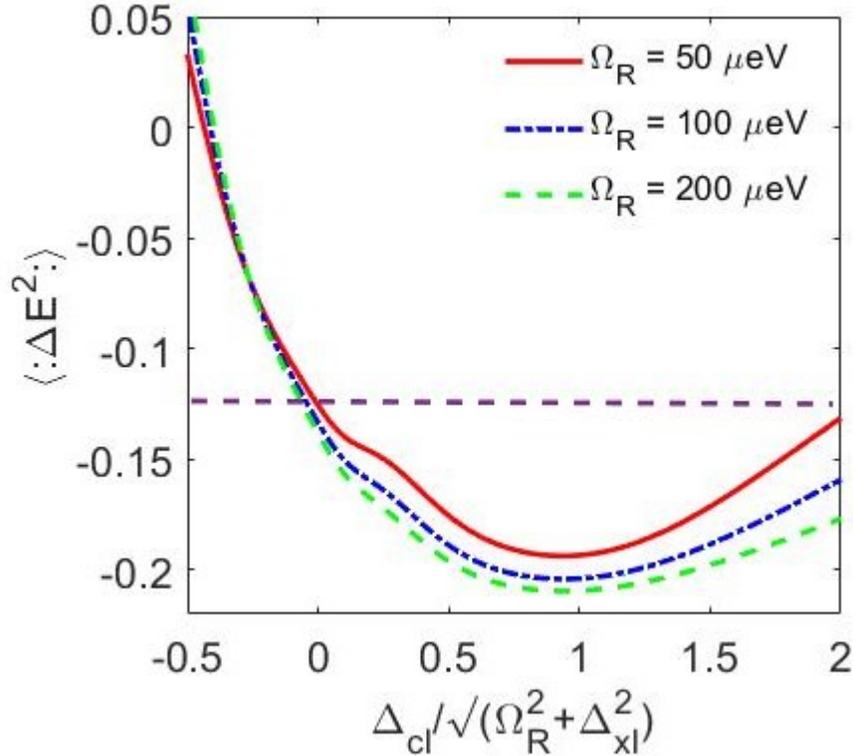

Fig. 2 (color online) Evolution of the normally ordered variance, $\langle :\Delta E^2: \rangle$, as a function of normalized cavity detuning, $\Delta_{cl}/\sqrt{\Omega_R^2 + \Delta_{xl}^2}$, for three different Rabi frequencies, $\Omega_R = 50\ \mu eV$ (solid red line), $100\ \mu eV$ (dashed-doted blue line), and $200\ \mu eV$ (dashed green line) without incorporating the exciton-phonon coupling. Magenta dashed line is shown as a reference representing the maximum achievable negative value of normally ordered variance, -0.125, in the fluorescence of an ideal two-level emitter in free space.

We have chosen the values of cavity coupling strength and cavity decay rate as $0.6\Omega_R$ and $0.9\Omega_R$, respectively throughout this paper. It can clearly be observed from Fig. 2 that the variance $\langle :\Delta E^2: \rangle$ goes to minimum at about $\Delta_{cl} \approx \sqrt{\Omega_R^2 + \Delta_{xl}^2}$. These values of minima are much below the maximum achievable negative value of variance, -0.125, of an ideal two-level emitter in free space as compared in the figure. Furthermore, the minimum value of variance, $\langle :\Delta E^2: \rangle = -0.214$, is achieved for $\Omega_R = 200\ \mu eV$. This obtained negative value is slightly lower than the reported value of variance -0.236 in the Ref. [38]. This is primarily due to the inclusion of finite value for pure dephasing ($\gamma'$) in our numerical calculations.

We now investigate the actually achievable variance or squeezing in the driven QDs, by considering the unavoidable exciton-phonon coupling as in Fig. 3. We show the evolution of variance, $\langle :\Delta E^2: \rangle$, as a function of normalized cavity detuning, $\Delta_{cl}/\sqrt{\Omega_R^2 + \Delta_{xl}^2}$, by incorporating the exciton-phonon coupling induced incoherent rates [see Eq. 5] at phonon bath temperature, $T = 4\ K$. It is clear from Fig. 3(a) that (i) the minima of variance occurs around $\Delta_{cl} \approx \sqrt{\Omega_R^2 + \Delta_{xl}^2}$ similar to Fig. 2 and (ii) order of minima are reversed as compared to Fig. 2, i.e., at -0.173 for $\Omega_R = 50\ \mu eV$ and at -0.122 for $\Omega_R = 200\ \mu eV$. These are quite in contrast and opposite to those obtained without incorporating the exciton-phonon coupling. Furthermore, it is also clear from Fig. 3 (a) that the deteriorating impact of exciton-phonon coupling on variance increases with the increasing Rabi frequency, causing eventually the shifting of variance for $\Omega_R = 200\ \mu eV$ above the minimum achievable value of variance of an ideal two-level emitter in free space. This Rabi frequency dependent effect of exciton-phonon coupling on variance can easily be understood from the analytically driven expressions of phonon-induced incoherent rates as a function of Rabi frequency [see Eq. 5]. In Fig. 3 (b), we show the evolution of variance for the same set of Rabi frequency and phonon bath temperature, $T$, but with exactly opposite exciton-laser detuning, $\Delta_{xl} = -\Omega_R$. It can be observed from Fig 3(b) that maximum negative value of variance -0.175 is obtained around $\Delta_{cl} \approx -\sqrt{\Omega_R^2 + \Delta_{xl}^2}$ for $\Omega_R = 50\ \mu eV$, while the obtainable value of variance around the same cavity detuning is now -0.145 for $\Omega_R = 200\ \mu eV$, which is clearly below the minimum achievable value of variance of an ideal two-level emitter in free space. Therefore, with the reversed exciton-laser detuning, $\Delta_{xl} = -\Omega_R$, the negativity of variance increases i.e. squeezing increases appreciably, particularly for $\Omega_R = 200\ \mu eV$ in comparison with that of $\Delta_{xl} = \Omega_R$. This observed enhancement in the squeezing at $\Delta_{xl} = -\Omega_R$ can be attributed to the asymmetry of phonon-induced rates, particularly, $\Gamma_{ph}^{\sigma^+ a}$ and $\Gamma_{ph}^{a^\dagger \sigma^-}$, with respect to the cavity-exciton detuning, $\Delta_{cx}$, because they are exactly four-fold greater than $\Gamma_{ph}^{\sigma^-}$ and $\Gamma_{ph}^{\sigma^+}$ [see Eq. 5]. For $\Delta_{xl}, \Delta_{cl} = +ve$ as employed in Fig. 3(a), the cavity-exciton detuning, $\Delta_{cx}$, also turns out to be positive. Now, for $\Delta_{cx} = +ve$ values, the phonon-induced rate, $\Gamma_{ph}^{\sigma^+ a}$ takes appreciably greater value as compared to $\Gamma_{ph}^{a^\dagger \sigma^-}$ particularly for $\Omega_R = 200\ \mu eV$ as can be understood from Fig. 1(b). This enables the dominant flow of population into the exciton state resulting in the greater steady state population in the exciton state. Due to the increased population in the exciton state, the negative value of variance shifts towards zero, i.e. squeezing also decreases. However, for $\Delta_{xl}, \Delta_{cl} = -ve$ as employed in Fig. 3(b), the cavity-exciton detuning, $\Delta_{cx}$, also turns out to be negative. For $\Delta_{cx} = -ve$ values, the dominant flow of population is exactly in opposite direction. Specifically, $\Gamma_{ph}^{a^\dagger \sigma^-}$ takes appreciably greater values compared to $\Gamma_{ph}^{\sigma^+ a}$ particularly for $\Omega_R = 200\ \mu eV$ as again can be understood from Fig. 1(b). Consequently, there is a phonon-assisted decrease in the exciton state population, which results the increase in the negative value of variance indicating the greater squeezing.

From the above discussion, it should be clear now that, for the dot-cavity parameters considered in this paper, the inclusion of unavoidable exciton-phonon coupling significantly decreases the negative value of the variance and hence the obtainable squeezing in comparison with that of without exciton-phonon coupling. However, the value of variance of -0.175 obtained for $\Delta_{xl}, \Delta_{cl} = -ve$ with $\Omega_R = 50\ \mu eV$ is still much below the minimum achievable value of variance, -0.125, in the fluorescence of an ideal two-level emitter in free space. Moreover, this obtained value of variance (squeezing) -0.175 (5.2 dB) is at least three-fold (four-fold) greater than the minimum value of variance (squeezing) -0.056 (1.1) obtained for a single QD without cavity coupling as reported in Refs. [28, 29]. Note that the phonon-induced enhancement in the negative value of variance of about -0.20 is reported in a strongly driven single QD without cavity coupling [37].

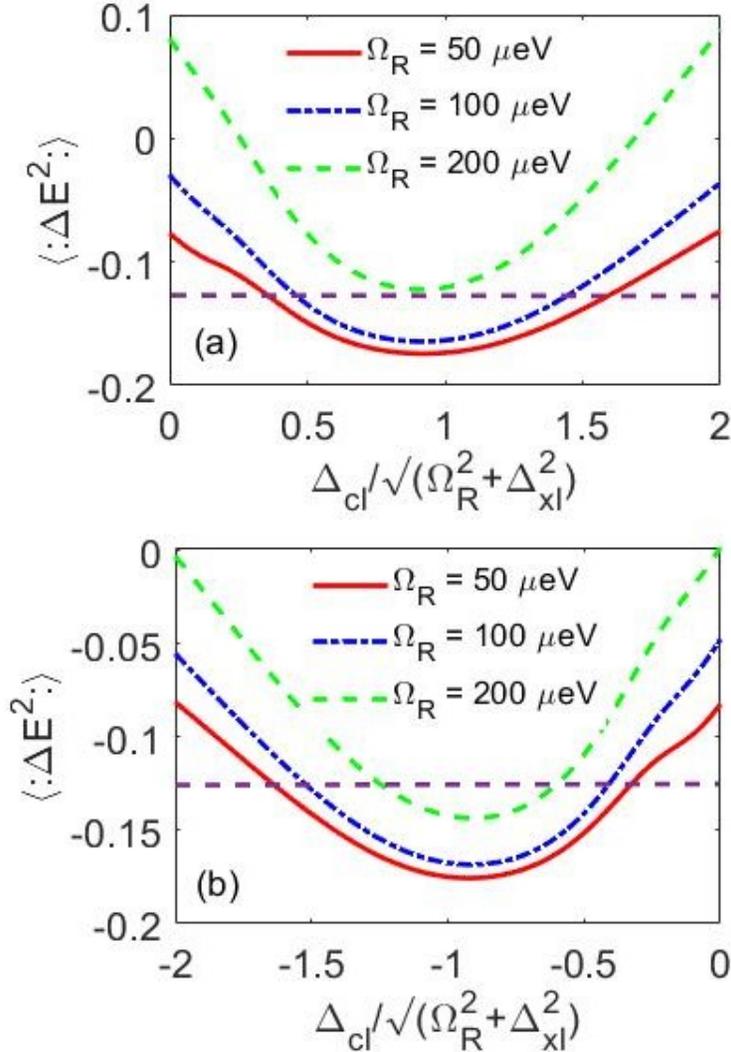

Fig. 3 (color online) Evolution of the normally ordered variance, $\langle:\Delta E^2:\rangle$, as a function of normalized cavity detuning, $\Delta_{cl}/\sqrt{\Omega_R^2 + \Delta_{xl}^2}$, for three different Rabi frequencies, $\Omega_R = 50\ \mu eV$ (solid red line), $100\ \mu eV$ (dashed-doted blue line), and $200\ \mu eV$ (dashed green line) with the inclusion of the exciton-phonon coupling for the two cases of exciton-laser detuning (a) $\Delta_{xl} = \Omega_R$ (b) $\Delta_{xl} = -\Omega_R$. Magenta dashed line is shown as a reference representing the maximum achievable negative value of normally ordered variance, -0.125, in the fluorescence of an ideal two-level emitter in free space.

However, in contrast, we employ a cavity coupling for enhancing the squeezing and deliberately chose the Rabi frequency to be in $\mu eV$ regime, primarily because of the realistic two practical reasons. Firstly, for example, the value of required laser power will be $12.7\ mW$ for a typical beam spot area of $100\ \mu m^2$ ($I = 1.27 \times 10^8\ W/m^2$) for our chosen Rabi frequency, $\Omega = \frac{pE}{\hbar} = 200\ \mu eV$, and for a typical value of electric dipole moment of $p = 9.7 \times 10^{-29}\ Cm$, of exciton state [46]. This can be easily afforded by a broad set of readily available CW lasers. In contrast, the required laser power of $24\,W (I = 2.4 \times 10^{11}\,W/m^2)$ for their chosen maximum Rabi frequency, $\Omega = 8\,meV$, is very difficult to achieve with the readily available CW lasers [47]. Secondly, for the experimental studies of CW laser driven QDs performed in $\mu eV$ regime, the effective polaron master equation captures even the effect of exciton-phonon coupling quite accurately [32, 48, 49].

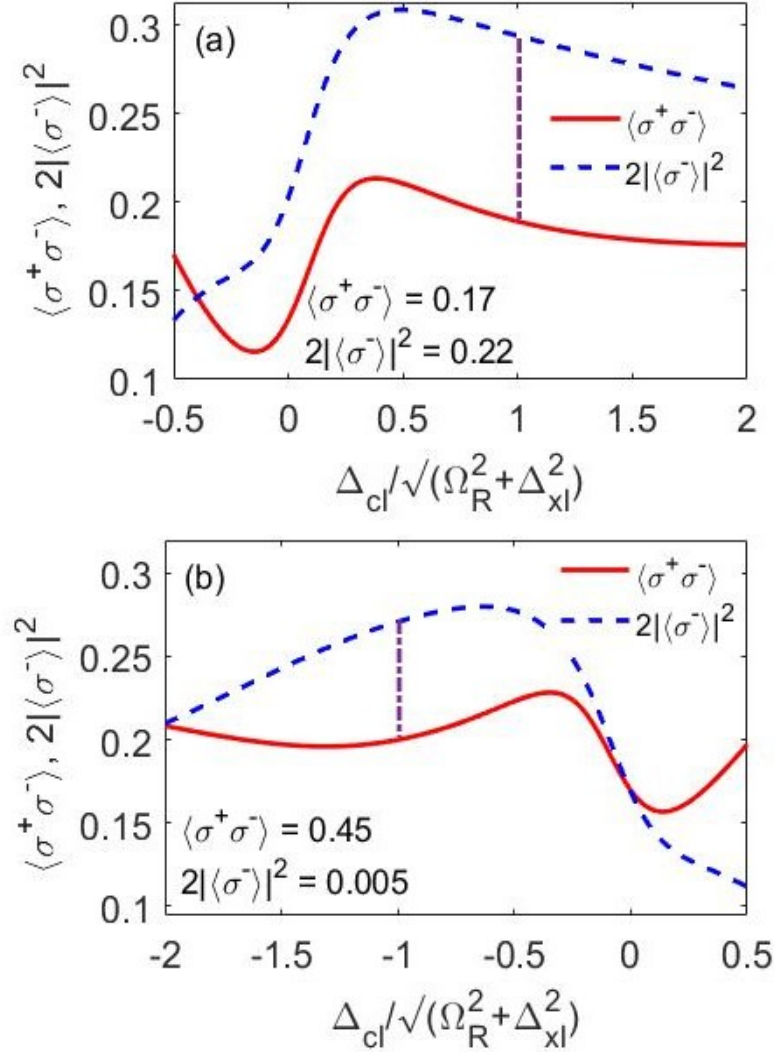

Fig. 4 (color online) Evolution of exciton state population, $\langle \sigma^+\sigma^-\rangle$, and coherence, $2|\langle \sigma^-\rangle|^2$, as a function of cavity detuning, $\Delta_{cl}/\sqrt{\Omega_R^2 + \Delta_{xl}^2}$, for the fixed value of Rabi frequency, $\Omega_R = 200\,\mu eV$, cavity coupling strength, $g_R = 0.6\Omega_R$, and cavity decay rate, $\kappa = 0.9\Omega_R$ for the two cases of (a) without exciton-phonon coupling at $\Delta_{xl} = \Omega_R$, (b) with exciton-phonon coupling at phonon-bath temperature, $T = 4K$ and $\Delta_{xl} = -\Omega_R$. The values of the exciton state population, $\langle \sigma^+\sigma^-\rangle$, and coherence, $2|\langle \sigma^-\rangle|^2$ indicated in the figure represent those values of without cavity coupling.

We have determined the maximally possible negative values of variance around $\Delta_{cl} = \sqrt{\Omega_R^2 + \Delta_{xl}^2}$ (from Fig. 2 and 3) irrespective of the exclusion or inclusion of exciton-phonon coupling. We further investigate the evolution of few more parameters such as exciton state population, $\langle \sigma^+\sigma^-\rangle$, and coherence, $2|\langle \sigma^-\rangle|^2$, as a function of cavity-laser detuning, $\Delta_{cl}$ as shown in Fig. 4 for gaining more insight and a better understanding. We also provide their values with the same set of parameters for the case of without cavity coupling in the figure for the purpose of explicit comparison. It should be clear from Fig. 4 (a) and 4(b) that the value of coherence, $2|\langle\sigma^-\rangle|^2$, gets enhanced significantly around cavity-laser detuning, $\Delta_{cl}= \sqrt{\Omega_R^2 + \Delta_{xl}^2}$, due to cavity coupling as compared to those without cavity coupling. However, it may be noted that in Fig. 4 (b) the value of coherence, $2|\langle\sigma^-\rangle|^2$ around $\Delta_{cl} = \sqrt{\Omega_R^2 + \Delta_{xl}^2}$ is quite less than that in Fig. 4(a), causing the decreased negative value of variance. This is primarily because of the partial loss of cavity-induced coherence due to the phonon-induced incoherent rates. Furthermore, the difference $[\langle\sigma^+\sigma^-\rangle - 2|\langle\sigma^-\rangle|^2]$ showing the maximum negative value around $\Delta_{cl}= \sqrt{\Omega_R^2 + \Delta_{xl}^2}$ is indicated in the figure by the vertical magenta dashed-dotted line. The significance of this difference is that it is directly proportional to the normally ordered variance $\langle:\Delta E^2:\rangle$ [see eqn. 6] which is the indicative of the degree of squeezing. Therefore, the enhancement in the squeezing can clearly be attributed to the cavity-enhanced coherence, $2|\langle\sigma^-\rangle|^2$, and reduced exciton state population around cavity-laser detuning, $\Delta_{cl}= \sqrt{\Omega_R^2 + \Delta_{xl}^2}$ for both with and without exciton-phonon coupling.

C. **Evolution of the Variance as a Function of Natural Decay and Pure Dephasing Rates**

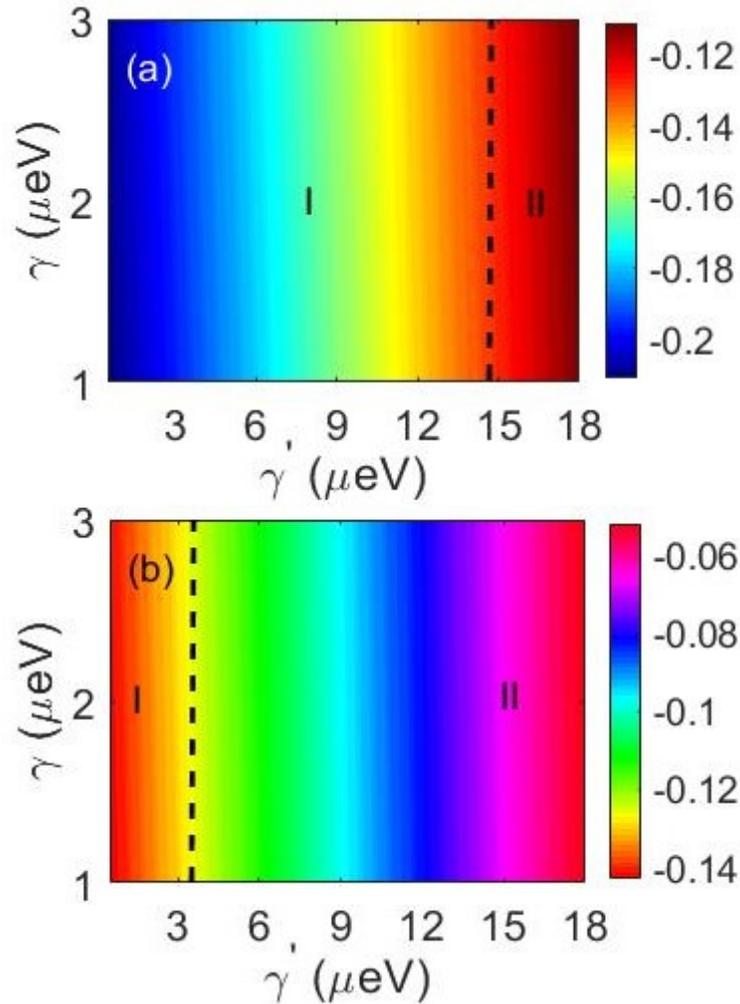

Fig. 5 (color online) Evolution of the normally ordered variance, $\langle:\Delta E^2:\rangle$, as a function of exciton state natural decay $(\gamma)$ and dephasing $(\gamma')$ rates, for a fixed value of Rabi frequency, $\Omega_R = 200 \,\mu eV$, cavity coupling strength, $g_R = 0.6\Omega_R$, and cavity decay rate, $\kappa = 0.9\Omega_R$ for the two cases of (a) without exciton-phonon coupling at $\Delta_{xl}= \Omega_R$, (b) with exciton-phonon coupling at phonon-bath temperature, $T = 4K$ and $\Delta_{xl}= -\Omega_R$.

Finally, we show the evolution of normally ordered variance $\langle :\Delta E^2: \rangle$ as a function of natural decay rate, $\gamma$, and pure dephasing rate, $\gamma'$, without and with exciton-phonon coupling in Fig. 5(a) and 5(b) respectively. It can clearly be observed from Fig 5(a) that the variance depicts quite weak dependence on radiative decay rate $\gamma$ and a quite stronger dependence on pure dephasing rate $\gamma'$. However, the substantial amount of negative variance still persists even when the pure dephasing rate, $\gamma'$, becomes significantly larger than the natural decay rate, $\gamma$. More specifically, the region I as indicated in the figure ($\gamma: 1-3, \gamma': 0.5-14$) has the value of variance well below -0.125 and similarly region II ($\gamma: 1-3, \gamma': 14-18$) possesses the value of variance well above -0.125. Now, in Fig 5(b), we analyze the evolution of variance with same set of parameters as those in Fig 5(a) with exciton-phonon coupling. It can clearly be understood that the variance again depicts a quite weak dependence on radiative decay rate $\gamma$ but a much stronger dependence on pure dephasing rate, $\gamma'$. Specifically, with exciton-phonon coupling the range of region I reduces drastically from $\gamma: 1-3, \gamma': 0.5-14$ to $\gamma: 1-3, \gamma': 0.5-4$, while the range of region II expands from $\gamma: 1-3, \gamma': 14-18$ to $\gamma: 1-3, \gamma': 4-18$. Nevertheless, the negativity of variance persists even when the pure dephasing rate, $\gamma'$, becomes quite larger than the natural decay rate, $\gamma$. This is clearly in sharp contrast and advantageous compared to the results reported in the bare QD where negativity does not persist at all in regime where pure dephasing rate, $\gamma'$, becomes comparable or greater than the natural decay rate, $\gamma$ [28]. Evidently, the persistence of the negativity of variance is solely due to the cavity-enhanced coherence. The smaller negative values of variance than those without exciton-phonon coupling in Fig. 5(a) can be attributed to the partial loss of cavity-enhanced coherence due to the phonon-induced incoherent rates as shown and described in Fig.4.

## IV. CONCLUSION

We have demonstrated the cavity-enhanced and decoherence immune squeezing in the resonance fluorescence of a cavity coupled single quantum dot using a rigorous theoretical formalism based on Polaron master equation theory. Compared to that of an ideal single two-level system in free space, we show a significant enhancement of squeezing due to cavity coupling of the QD even in the presence of deteriorating exciton-phonon interaction. Furthermore, we have also shown that a four-fold enhancement in squeezing can be achieved in a cavity coupled single QD in contrast with that of an isolated one. The squeezing is found to be fairly robust for the increment in pure dephasing and persists even when pure dephasing rate exceeds the radiative decay rate. We have shown that the enhancement in squeezing is mainly facilitated by the cavity-enhanced coherence for an appropriate amount of cavity detuning. The phonon-bath temperature dependent incoherent effects are shown to be detrimental and do reduce the squeezing. However, the deteriorating effects of phonon-induced incoherent rates on squeezing can be partially circumvented by properly adjusting the detunings.

**ACKNOWLEDGEMENT**

This work is partially supported by a Dr. D. S. Kothari Postdoctoral Research Fellowship, University Grant Commission, India, through Grant No.F.4-2/2006 (BSR)/PH/15-16/0077.

**REFERENCES**
[1] D. Kleppner, Inhibited Spontaneous Emission**,** Phys. Rev. Lett. **47**, 233 (1981).
[2] P. Lodahl, A. F. van Driel, I. S. Nikolaev, A. Irman, K. Overgaag, D. Vanmaekelbergh, andW. L. Vos, Controlling the dynamics of spontaneous emission from quantum dots by photonic crystals, Nature, **430**, 654 (2004).
[3] H. Toida, T. Nakajima, and S. Komiyama, Vacuum Rabi splitting in a semiconductor circuit QED System, Phys. Rev. Lett. **110**, 066802 (2013).

[4] B. R. Mollow, Power spectrum of light scattered by two-level systems, Phys. Rev. **188**, 1969 (1969).
[5] H. J. Kimble, M. Dagenais, and L. Mandel, Photon antibunching in resonance fluorescence, Phys. Rev. Lett. **39**, 691 (1977).
[6] M. A. Nielsen and I. L. Chuang, *Quantum Computation and Quantum Information* (Cambridge University Press, Cambridge, 2000).
[7] H. J. Kimble, The quantum internet, Nature (London) **453**, 1023 (2008).
[8] E. Knill, R. Laflamme, and G. J. Milburn, A scheme for efficient quantum computation with linear optics, Nature (London) **409**, 46 (2001).
[9] T. D. Ladd, F. Jelezko, R. Laflamme, Y. Nakamura, C. Monroe, and J. L. O'Brien, Quantum computers, Nature (London) **464**, 45 (2010).
[10] W. Wang, Y. Wu, Y. Ma, W. Cai, L. Hu, X. Mu, Y. Xu, Zi-Jie Chen, H. Wang, Y.P. Song, H. Yuan, C.-L. Zou, L.-M.Duan, and L. Sun, Heisenberg-limited single-mode quantum metrology in a superconducting circuit, Nature Comm. **10**, 4382 (2019).
[11] K. Liu, C. Cai, J. Li, L. Ma, H. Sun, and J. Gao, Squeezing-enhanced rotating-angle measurement beyond the quantum limit, Appl. Phys. Lett. **113**, 261103 (2018).
[12] L. F. Buchmann, S. Schreppler, J. Kohler, N. Spethmann, and D. M. Stamper-Kurn, Complex squeezing and force measurement beyond the standard quantum limit, Phys. Rev. Lett. **117**, 030801 (2016).
[13] J. B. Clark, F. Lecocq, R. W. Simmonds, J. Aumentado and J. D. Teufel, Sideband cooling beyond the quantum backaction limit with squeezed light, *Nature* **541**, 191 (2017).
[14] N. Német and S. Parkins, Enhanced optical squeezing from a degenerate parametric amplifier via time-delayed coherent feedback, Phys. Rev. A **94**, 023809 (2016).
[15] H. H. Adamyan, J. A. Bergou, N. T. Gevorgyan, and G. Y. Kryuchkyan, Strong squeezing in periodically modulated optical parametric oscillators, Phys. Rev. A **92**, 053818 (2015).
[16] J. F. Corney, J. Heersink, R. Dong, V. Josse, P. D. Drummond, G. Leuchs, and U. L. Andersen, Phys. Rev. A **78**, 023831 (2008).
[17] R. E. Slusher, L. W. Hollberg, B. Yurke, J. C. Mertz, and J. F. Valley, Observation of squeezed states generated by four-wave mixing in an optical cavity, Phys. Rev. Lett. 55, 2409 (1985).
[18] C. F. McCormick, A. M. Marino, V. Boyer, and P. D. Lett, Strong low-frequency quantum correlations from a four-wave-mixing amplifier, Phys. Rev. A 78, 043816 (2008).
[19] Q. Glorieux, R. Dubessy, S. Guibal, L. Guidoni, J.-P. Likforman, T. Coudreau, and E. Arimondo, Double- microscopic model for entangled light generation by four-wave mixing, Phys. Rev. A **82**, 033819 (2010).
[20] Jia-pei Zhu, Hui Huang, and Gao-xiang Li, Phonon-mediated squeezing of the cavity field off-resonantly coupled with a coherently driven quantum dot, J. Appl. Phys. **115**, 033102 (2014).
[21] A. Ourjoumtsev, A. Kubanek, M. Koch, C. Sames, P. Pinkse, G. Rempe, K. Murr, Observation of squeezed light from one atom excited with two photons, Nature **474**, 623 (2011).
[22] J. Estève, C. Gross, A. Weller, S. Giovanazzi, and M. K. Oberthaler, Squeezing and entanglement in a Bose-Einstein condensate, Nature (London) **455**, 1216 (2008).
[23] M. Rashid, T. Tufarelli, J. Bateman, J. Vovrosh, D. Hempston, M. S. Kim, and H. Ulbricht, Experimental realization of a thermal squeezed state of levitated optomechanics, Phys. Rev. Lett.**117**, 273601 (2016).
[24] V. Peano, H. G. L. Schwefel, Ch. Marquardt, and F. Marquardt, Intracavity squeezing can enhance quantum-limited optomechanical position detection through deamplification, Phys. Rev. Lett. **115**, 243603 (2015).

[25] K. Qu and G. S. Agarwal, Generating quadrature squeezed light with dissipative optomechanical coupling, Phys. Rev. A **91**, 063815 (2015).

[26] A. K. Chauhan and A. Biswas, Atom-assisted quadrature squeezing of a mechanical oscillator inside a dispersive cavity, Phys. Rev. A **94**, 023831 (2016).

[27] D. F. Walls and P. Zoller, Reduced quantum fluctuations in resonance fluorescence, Phys. Rev. Lett. **47**, 709 (1981).

[28] P. Kumar and A. G. Vedeshwar, Generation of frequency-tunable squeezed single photons from a single quantum dot, J. Opt. Soc. Am. B **35,** 3035 (2018).

[29] C. H. H. Schulte, Jack Hansom, Alex E. Jones, Clemens Matthiesen, C. L. Gall, and Mete Atatüre, Quadrature squeezed photons from a two-level system, Nature Lett. **525**, 222 (2015).

[30] M. Glässl, A. M. Barth, and V. M. Axt, Proposed robust and high-fidelity preparation of excitons and biexcitons in semiconductor quantum dots making active use of phonons, Phys. Rev. Lett. **110**, 147401 (2013).

[31] J. H. Quilter, A. J. Brash, F. Liu, M. Glässl, A. M. Barth, V. M. Axt, A. J. Ramsay, M. S. Skolnick, and A. M. Fox, Phonon-assisted population inversion of a single InGaAs/GaAs quantum dot by pulsed laser excitation, Phys. Rev. Lett. **114**, 137401 (2015).

[32] S. Hughes and H. J. Carmichael, Phonon-mediated population inversion in a semiconductor quantum-dot cavity system, New J. Phys. **15**, 053039 (2013).

[33] H. Kim, T. C. Shen, K. R. Choudhury, G. S. Solomon, and E. Waks, Resonant interactions between a Mollow triplet sideband and a strongly coupled cavity, Phys. Rev. Lett. **113**, 027403 (2014).

[34] Y.-J. Wei, Y. He, Y.-M.He, C.-Y.Lu, J.-W. Pan, C. Schneider, M. Kamp, S. Höfling, D. P. S. McCutcheon, and A. Nazir, Temperature-dependent Mollow triplet spectra from a single quantum dot: Rabi frequency renormalization and sideband linewidth Insensitivity, Phys. Rev. Lett. **113**, 097401 (2014).

[35] C. Roy and S. Hughes, Polaron master equation theory of the quantum-dot Mollow triplet in a semiconductor cavity-QED system, Phys. Rev. B **85**, 115309 (2012).

[36] P. Kumar and A. G. Vedeshwar, Phonon-assisted control of the single-photon spectral characteristics in a semiconductor quantum dot using a single laser pulse, Phys. Rev. A **96**, 033808 (2017).

[37] J. I.-Smith, A. Nazir, and D. P.S. McCutcheon, Vibrational enhancement of quadrature squeezing and phase sensitivity in resonance fluorescence, Nat. Comm. **10**, 3034 (2019).

[38] P. Grünwald and W. Vogel, Optimal squeezing in resonance fluorescence via atomic-state purification, Phys. Rev. Lett. **109**, 013601 (2012).

[39] G. D. Mahan, *Many-Particle Physics* (Plenum, New York, 1990).

[40] I. Wilson-Rae and A. Imamoğlu, Quantum dot cavity-QED in the presence of strong electron-phonon interactions Phys. Rev. B **65**, 235311 (2002).

[41] D. P. S. McCutcheon and A. Nazir, Model of the optical emission of a driven semiconductor quantum dot: phonon-enhanced coherent scattering and off-resonant Sideband Narrowing, Phys. Rev. Lett. **110**, 217401 (2013).

[42] N. Makri and D. E. Makarov, Tensor propagator for iterative quantum time evolution of reduced density matrices. I. Theory, J. Chem. Phys. **102**, 4600 (1995).

[43] C. Roy and S. Hughes, Influence of electron–acoustic-phonon scattering on intensity power broadening in a coherently driven quantum-dot–cavity system, Phys. Rev. X **1**, 021009 (2011).

[44] D. P. S. McCutcheon and A. Nazir, Quantum dot Rabi rotations beyond the weak exciton–phonon coupling regime**,** New J. Phys. **12**, 113042 (2010).

[45] H.-P. Breuer and F. Petruccione, *The Theory of Open Quantum Systems* (Oxford University Press, Oxford, 2002).

[46] S. Stobbe J. Johansen, I. S. Nikolaev,; T. Lund-Hansen, P. T. Kristensen, J. M. Hvam, W. L. Vos, P. Lodahl, Accurate measurement of the transition dipole moment of self-assembled quantum dots, CLEOE-IQEC 2007, https://doi.org/10.1109/CLEOE-IQEC.2007.4387029.
[47] O. Svelto, Principles of Lasers, (Springer Science & Business Media, 2010).
[48] A. Ulhaq, S. Weiler, C. Roy, S. M. Ulrich, M. Jetter, S. Hughes, and P. Michler, Detuning-dependent Mollow triplet of a coherently-driven single quantum dot, Opt. Exp. **21**, 4382 (2013).
[49] S. Unsleber, S. Maier, Dara P. S. McCutcheon, Yu-Ming He, M. Dambach, M. Gschrey, N. Gregersen, J. Mørk, S. Reitzenstein, S. Höfling, C. Schneider, and M. Kamp, Observation of resonance fluorescence and the Mollow triplet from a coherently driven site-controlled quantum dot, Optica **2**, 1072, (2015).